\documentclass{article}

\usepackage{arxiv}

\usepackage[utf8]{inputenc} % allow utf-8 input
\usepackage[T1]{fontenc}    % use 8-bit T1 fonts
\usepackage{hyperref}       % hyperlinks
\usepackage{url}            % simple URL typesetting
\usepackage{booktabs}       % professional-quality tables
\usepackage{amsfonts}       % blackboard math symbols
\usepackage{nicefrac}       % compact symbols for 1/2, etc.
\usepackage{microtype}      % microtypography
\usepackage{lipsum}
\usepackage{changepage}
\usepackage{graphicx}
\usepackage{amsmath}
\usepackage{amsfonts}
\usepackage{amssymb}
\usepackage{makeidx}
\usepackage{algorithm,algorithmic}
\usepackage[switch]{lineno}
\usepackage{xcolor}
\usepackage{bm}
\usepackage[justification=centering]{caption}
\usepackage{amsfonts}
\usepackage[bbgreekl]{mathbbol}
\DeclareSymbolFontAlphabet{\mathbbm}{bbold}
\DeclareSymbolFontAlphabet{\mathbb}{AMSb}%
\usepackage{amsthm}
\newtheorem{theorem}{Theorem}
\newtheorem{problem}{Problem}
\newtheorem{definition}{Definition}

\newtheorem{lemma}{Lemma}
\newtheorem{assumption}{Assumption}
\title{Approximate Dynamic Programming based\\ Model Predictive Control of Nonlinear systems}

\author{Keerthi Chacko,
  Midhun T. Augustine, S. Janardhanan, Deepak U. Patil, and I. N. Kar\\Department of Electrical Engineering\\
  Indian Institute of Technology Delhi, India\\
  \texttt{keerthichacko@gmail.com} } 
\begin{document}
\maketitle
\large
\begin{abstract}
This paper studies the optimal control problem for discrete-time nonlinear systems and an approximate dynamic programming-based Model Predictive Control (MPC) scheme is proposed for minimizing a quadratic performance measure. In the proposed approach, the value function is approximated as a quadratic function for which the parametric matrix is computed using a switched system approximate of the nonlinear system. The approach is modified further using a multi-stage scheme to improve the control accuracy and an extension to incorporate state constraints. The MPC scheme is validated experimentally on a multi-tank system which is modeled as a third-order nonlinear system. The experimental results show the proposed  MPC scheme results in significantly lesser online computation compared to the Nonlinear MPC scheme.
\end{abstract}

% keywords can be removed

\section{Introduction}\label{sec1}

Model predictive control is an advanced control technique that can handle complex multivariable dynamics, respecting its constraints that might have a competing set of objectives \cite{kb:15}. Hence MPC is widely used in industrial systems that operate with strict performance criteria having limitations on state and control. %Linear MPC with its strong theoretical foundation is extensively used in industry. However, systems are inherently nonlinear. And designing a nonlinear MPC is challenging.
% Disadvantages of MPC and motivation for the use of DP
MPC involves periodic online optimization along with a receding horizon implementation of the control law. In MPC, the optimization problem is normally solved using either 
 nonlinear programming (NLP) or dynamic programming (DP) approaches. The NLP is based on iterative algorithms in which the elements of the decision vector
are optimized together, whereas DP is a recursive approach in which the elements of the decision vector
are optimized recursively, i.e., one at a time.
 The computational complexity of both NLP and DP increases with the prediction horizon which limits the application of MPC to a wide range of systems. Techniques have been devised to solve the issue of intractable optimal policy by an approximate policy. Approximate dynamic programming (ADP) is a solution to the computational complexity of the DP approach\cite{jml:05}.
%This motivates us to explore the possibility of using Approximate Dynamic Programming (ADP) with MPC.
%\par Dynamic programming is a sequential backward in time algorithmic method to get the optimal solutions with offline calculations. It provides the most efficient and reliable control for complex systems.
% DP TO ADP
%Dynamic programming suffers from the "curse of dimensionality". This issue is addressed to a large extent by the use of Approximate Dynamic Programming (ADP)\cite{jl:05,jl:04}. 
ADP is an optimization algorithm that aims at computational tractability for complex systems by efficiently approximating the value function in DP.  %ADP targets the most accurate solutions respecting computational constraints. 
ADP is used to compute the near-optimal control actions using the approximate value function \cite{fl:12}.
%Approximation of value functions or policies has gained a lot of attention recently \cite{ld:19}.
ADP can be categorized into  the following classes based on the type of approximation:
\begin{enumerate}
    \item Policy approximation: which is based on approximation in the policy (feedback control) space.
%  \item Cost function approximation
    \item Value approximation:  which is based on approximation in the value function (optimal cost-to-go function). In this paper, we focus on the value function approximation.
    %Value approximation: which is based on the approximation of the optimal cost-to-go function or the value function. In this paper, we focus on the value function approximation.
    \item Direct look ahead: is based on the direct enumeration with receding horizon implementation. It can be either a single-step look ahead or a multi-step look ahead. This approach can be combined with both the policy and value approximation to reduce the computation further.
\end{enumerate}
\par ADP enables DP to be applied to a wider range of applications that require the solution of a large-scale dynamic system \cite{pr:21,adk:21,sdh:21}. Recently, ADP has emerged as a promising technique and is now used for a wide variety of applications which includes battery storage applications \cite{nz:20}, robotic surveillance \cite{mp:16}, mechanical systems \cite{ts:21}, etc. The ADP along with neural network identification of unknown dynamics is proposed \cite{ld:19}, where the neural network identification is used to relax the knowledge of the system dynamics. A multi-step heuristic ADP algorithm \cite{bl:19} has a discounted performance index to convert the problem to a regulation problem. This algorithm uses a neural network-based actor-critic approach for implementation. %Most of the proposed techniques involving ADP were based on simulations 
A kernel-based ADP \cite{xs:14}  is derived and applied on an inverted pendulum which is based on policy approximation. The convergence analysis of DP for a nonlinear system is detailed for an event-based control \cite{hz:19}. %But event-based control is not robust enough to handle control constraints and disturbances.  
Switched control strategies to minimize the cost function over an infinite time horizon are given \cite{wh:12} which proposes the idea of ADP in switched systems. %This aims at computationally efficient formulation but is based on a particular structure of discrete-time switched LQR structure and the choice of the tuning parameters is arbitrary.
ADP is used to solve resource allocation problems \cite{af:20} for both finite and infinite horizon problems, which focuses on the specific application of resource allocation that can be modeled as a stochastic problem. Another approach that discusses computationally tractability in a stochastic framework was recently proposed \cite{pn:22}. The connection between the MPC and ADP was established in \cite{dp:05,dp:12}. ADP based MPC of nonlinear systems is discussed \cite{lj:11}. While each of these approaches has its advantages and disadvantages, none of them utilize the switched system framework to approximate the value function. 
%The main focus of \cite{mg:21} is on stability for which they add a contracting factor in the cost function which is conservative and may result in feasibility in subsequent iterations.
% Disadvantage of ADP and the problem addressed in this paper
%\par The optimal control problem becomes intractable as the state and control dimensions increase. The time and memory requirements grow exponentially.  We are interested in an optimization algorithm tailored to bring down the online computational load. The stochastic framework requires a lot of computation. To overcome this requirement we use ADP for stochastic problems \cite{af:20}.
% Other Special issues for the process control problem
\par One of the major application areas of the MPC is the process control industries. Process control systems generally have a large number of state variables and control inputs with physical constraints\cite{KC2}. MPC has the ability to handle multi-variable systems and can incorporate input/state constraints which makes it a suitable choice for the process control industries \cite{jl:09,jl:04}.
In practice, most of the process control systems are nonlinear which makes the MPC optimization problem nonlinear and can be non-convex. Online learning and optimization is costly as we cannot operate the process at random points to explore the state and control space\cite{mm:20}. Although a large number of simulations can be carried out offline, the dynamics of the process control system are quite complex that it limits the exploration of state space offline \cite{jl:04}. All these factors render the application of MPC for process control systems a challenging task. %The multi-tank system discussed here is an under-actuated system with only one control input to maintain the levels of multiple tanks. Calculating an appropriate control input that carries out the task of maintaining the levels of the tank is a challenging task.

% Literature on convergence and what it lacks
%In this paper we have derived the convergence analysis for the tracking control for a nonlinear system. The effect of single-stage and multi-stage ADP on convergence is also discussed.

% Main features of the proposed technique

\par In this paper, we discuss the design and analysis of an ADP based MPC scheme and its validation in a multi-tank system. The proposed approach is based on a direct look ahead with value function approximation using a switched system model of the nonlinear system. Firstly, the paper discusses the proposal of single-stage ADP based MPC and its issues. Then move on to discuss the multi-stage ADP based MPC which overcomes the issues caused by single-stage ADP and improves the performance of the MPC algorithm significantly.
The MPC scheme is implemented in a multi-tank system \cite{KC1,ind:03} which serves as a benchmark system to test control algorithms.  Our objective is to maintain the levels of the tanks by controlling the speed of the motor. The variable cross-section of the tanks contributes to the non-linearity in the system rendering the level control a challenging task. Other factors which contribute to the non-linearity are the flow dynamics and valve characteristics. The control algorithms are tested in this setup to verify their efficiency.
In the final section, we suggest the method to handle state constraints and a comparison of all the proposed methods with nonlinear model predictive control. We believe, this work is the first attempt to combine the concept of switched dynamics to efficiently approximate value functions and apply them to multi-tank dynamic optimization problems. 
% Features of Experimental setup
% Contributions
The primary contributions of this paper focusing on the design of ADP based MPC are summarized below: \begin{enumerate}
    \item \textbf{ADP-based MPC using switched system model}: In the proposed MPC strategy, the value function from next instant  
   is approximated as a quadratic function using a switched affine system (SAS) model of the nonlinear system. 
   Using the SAS model, the parameter matrices which approximates the value function are computed offline. This reduces the online computation required for the proposed scheme. Further, we modify the MPC algorithm using a multi-stage scheme   to improve the control accuracy.
   % The inclusion of the ADP strategy could significantly reduce the computational load of the MPC controller.
    \item \textbf{Extension of the MPC scheme to incorporate state constraints}: This paper  focuses on the MPC with an unconstrained state case, i.e., only the control input is constrained. However, in Section 5, the ADP-based MPC scheme is extended to nonlinear systems with both input and state constraints where the state constraints are considered bounded. 
   % \item A posteriori analysis that establishes the stability of the proposed techniques.
    \item \textbf{Experimental validation}: Experimental study carried out on a multi-tank system to prove the efficacy of the proposed technique. The performance of the ADP based MPC scheme is compared with Nonlinear MPC scheme in terms of online computation time and integral square error which illustrated the effectiveness of the proposed scheme in terms of computation reduction.  
\end{enumerate}

% Organization

%\par The paper is organized as follows: problem formulation is given in section II. Section III explains the preliminaries on dynamic programming and approximate dynamic programming. Explanation of the proposed scheme of approximated dynamic programming based MPC is explained in section IV. MPC with bounded state constraints is explained in section V. Experimental validation and analysis are presented in Section VI. Section VII concludes this paper.

\subsection{Notations}

The set of all positive integers, real numbers, natural numbers and the n-dimensional Euclidean space are denoted by $\mathbb{Z}_{+}, \mathbb{R}_+, \mathbb{N},$ and $\mathbb{R}^n$ respectively. The cardinality of the set $\mathbb{S}$ is denoted as $|\mathbb{S}|$. The Euclidean norm with weight Q is represented as $\left\lVert{x}\right\rVert_Q$:$\sqrt{x^T Q x}$. The notation $x_{l|k}$ represents the state at $l-{th}$ instant predicted using the model, given the state at $k-{th}$ instant. ${P}>0$ $ ({P}\geq 0)$ denotes ${P}$ is a real symmetric positive definite (semidefinite) matrix.  Finally, ${I}_{n}$ represents the identity matrix of order $n \times n$.

%$x_N$ represent the state at $N^{th}$ instant.
%$x_{k}$ denotes the state at $k^{th}$ instant. %Vectors and matrices are represented by lower and upper case letters respectively. 
%The n-dimensional real (Euclidean) space is denoted by $\mathbb{R}^n$. The set of all natural numbers is denoted by $\mathbb{N}$ and the set of all positive integers and real numbers are denoted by $\mathbb{Z}_{+},\mathbb{R}_+$. The Q-weighted Euclidean norm is denoted as $\left\lVert{x}\right\rVert_Q$:$\sqrt{x^T Q x}$ and $|\mathbb{S}|$ denotes the cardinality of the set $\mathbb{S}$. The notation $x_{l|k}$ represents the state at $l^{th}$ instant predicted using the model, given the state at $k^{th}$ instant. ${P}>0$ $ ({P}\geq 0)$ denotes ${P}$ is a real symmetric positive definite (semidefinite) matrix.  Finally, ${I}_{n}$ represents the identity matrix of order $n \times n$.

%The set of all permissible states and control is represented by $\mathbb{X}$ and $\mathbb{U}$ respectively (i.e. admissible states and control). 
% $V_{k}^{*}$ represents the optimal cost at $k^{th}$ instant and $\hat{V}_{k}$ represents the approximate cost at $k^{th}$ instant.

%%%%%%%%%%%%%%%%%%%%%%%%%% two line equation %%%%%%%%%%%%%%%%%%%%%%%%
%\begin{eqnarray}
%s(nT_{s}) &= &s(t)\times \sum\limits_{n=0}^{N-1} \delta (t-nT_{s}) %\xleftrightarrow{\mathrm{DFT}}  S \left(\frac{m}{NT_{s}}\right) %\nonumber\\
%&= &\frac{1}{N} \sum\limits_{n=0}^{N-1} \sum\limits_{k=-N/2}^{N/2-1} %s_{k} e^{\mathrm{j}2\pi k\Delta fnT_{s}} e^{-j\frac{2\pi}{N}mn}
%\end{eqnarray}
%%%%%%%%%%%%%%%%%%%%%%%%%% two line equation %%%%%%%%%%%%%%%%%%%%%%%%

\section{Problem Formulation}\label{sec2}
Consider the nonlinear system in discrete-time
\begin{equation}
 x_{k+1}=f(x_{k},u_{k})
\end{equation}
where $k \in \mathbb{Z}_{+}$ is the discrete time instant,
$x_{{k}}\in \mathbb{X} \subseteq \mathbb{R}^{n}$ is the state vector, $u_{k}\in \mathbb{U}\subseteq \mathbb{R}^{m}$ is the control input vector and $\mathbb{X},\mathbb{U}$ are the constraint sets for the  state and control vectors. To simplify the problem, we consider MPC of discrete-time nonlinear systems (1) for which the state is considered to be unconstrained (i.e., $\mathbb{X}=\mathbb{R}^{n}$) and the control input is constrained. The extension of the proposed approach to MPC with bounded state constraints (polytopic constraints) will be discussed in section 5. Let $x_r$ be the set-point given to the controller.
The control input sequence for the MPC at time instant $k$ is 
\begin{equation}
  U_{k}=(u_{k|k},u_{k+1|k},...,u_{k+N-1|k})  
\end{equation}
The cost function for the MPC scheme at time instant $k$ with prediction horizon $N$ is defined as
\begin{equation}
J_{N} (x_k,U_k)=x_{k+N|k}^{T}Q_{N}x_{k+N|k}+\sum_{l=k}^{k+N-1}x_{l|k}^{T}Qx_{l|k}+u_{l|k}^{T}Ru_{l|k}
\end{equation}
where $Q_{N} \geq 0, Q\geq 0, R>0$ are the weighting matrices. To simplify the notation, we denote $L(x_{l|k},u_{l|k})=x_{l|k}^{T}Qx_{l|k}+u_{l|k}^{T}Ru_{l|k}.$
 Now, the MPC problem for nonlinear system is defined as follows
 
\begin{problem}
For the nonlinear system (1) with the current state $x_{k}$ given, compute the control input sequence $U_{k}$ by solving the optimization problem
\begin{equation}
\begin{aligned}
    &\underset{U_{k}}{min} \hspace{.2cm}J_{N} (x_k,U_k)\\
subject\hspace{.1cm} to:\hspace{.1cm}  &U_{k} \in \mathbb{U}^{N}\\
 &x_{l+1|k}=f(x_{l|k},u_{l|k}), \hspace{.5cm}k \in \mathbb{Z}^{+},\hspace{.1cm} l=k,...,k+N-1.
 \end{aligned}
 \end{equation}

\end{problem}
In MPC the optimization problem (4) is solved at each instant and the first element of the control sequence: $u_{k|k}^{*}$ is applied to the system which results in a receding horizon scheme.
In the proposed MPC scheme, the control input $u_{k|k}^{*}$ is computed using an approximate dynamic programming method that uses a switched system model of the nonlinear system.
For a given operating point $(x_{o},u_{o}),$ the nonlinear system (1) can be linearized as 
\begin{equation}
    x_{k+1} = Ax_{k}+Bu_{k}
\end{equation}
%where  $A=\frac{\partial f} {\partial x_k} (x_o,u_o) $ and $ B=\frac{\partial f} {\partial u_k} (x_o,u_o).$ 
We denote $\mathbb{U}_{v} =\{v_{1},v_{2},...,v_{M}\}$ as the discrete (quantized) set of control inputs and $\mathbb{U}_{v}\subset \mathbb{U}.$ Now, the LTI system (5) with discrete set of controls can be represented as a switched affine system
\begin{equation}
    x_{k+1}=Ax_{k}+B_{\sigma_k}
\end{equation}
where $\sigma_k \in \{1,2,...,M\}$ is the switching index and $B_{\sigma_k}=Bv_{\sigma_k}.$ By defining the augmented state vector $\bar{x}=\left[\begin{matrix}
x\\1
\end{matrix}\right] \in \mathbb{R}^{n+1},$ the state equation (6) can be compactly represented as \cite{md:22}
\begin{equation}
\bar{x}_{k+1}=\bar{A}_{\sigma_k}\bar{x}_{k} 
\end{equation}
where $\bar{A}_{\sigma_k}=\left[\begin{matrix}
A& B_{\sigma_k}\\ 0& 1
\end{matrix}\right]\in \mathbb{R}^{(n+1) \times (n+1)}$. Similarly the cost function (3) can be represented using the augmented state as
\begin{equation}
J_{N}=\bar{x}_{k+N}\bar{Q}_{N}\bar{x}_{k+N}+\sum_{l=k}^{k+N-1}\bar{x}_{l|k}^{T}\bar{Q}_{\sigma_{l|k}}\bar{x}_{l|k}
\end{equation}
where $\bar{Q}_{\sigma_k}=\left[\begin{matrix}
Q& 0\\ 0& v_{\sigma_k}^{T}Rv_{\sigma_k}
\end{matrix}\right]\in \mathbb{R}^{(n+1) \times (n+1)}$ and $\bar{Q}_{N}=\left[\begin{matrix}
Q& 0\\ 0& 0
\end{matrix}\right] \in \mathbb{R}^{(n+1) \times (n+1)}.$

%%%%%%%%%%%%%%%%%%%%%%%%%%%% FIGURE %%%%%%%%%%%%%%%%%%%%%%%%%%%%%%%
%\begin{figure}[t]
%\centerline{\includegraphics[width=342pt,height=9pc,draft]{empty}}
%\caption{This is the sample figure caption.\label{fig1}}
%\end{figure}

%\begin{figure*}
%\centerline{\includegraphics[width=342pt,height=9pc,draft]{empty}}
%\caption{This is the sample figure caption.\label{fig2}}
%\end{figure*}
%%%%%%%%%%%%%%%%%%%%%%%%%%%% FIGURE %%%%%%%%%%%%%%%%%%%%%%%%%%%%%%%
\section{Preliminaries}
\subsection{The Dynamic Programming approach}
Dynamic programming solves the optimal control problem recursively using a  cost-to-go function $V_{k},$ where $V_{k}$ is the cost accumulated from the current instant till the end. For the dynamic programming-based MPC scheme, we denote the cost-to-go function $V_{l|k}$ which is the cost accumulated from the time instant $l$ till the instant $k+N.$
For the optimal cost function (2), $V_{l|k}$ is defined as
\begin{equation}
V_{l|k}=L(x_{l|k},u_{l|k})+V_{l+1|k}  \hspace{1cm} V_{k+N|k}=x_{k+N|k}^{T}Q_{N}x_{k+N|k}
\end{equation}
where $l=k+N-1,...,k+1,k.$ In dynamic programming, the optimal cost-to-go function (value function) and the optimal control input are computed recursively by solving
	\begin{equation}
	\begin{aligned}
&V_{l|k}^{*}= \underset{u_{l|k}\in \mathbb{U}}{min} \hspace{.2cm} L(x_{l|k},u_{l|k})+ V_{l+1|k}^{*}(x_{l+1|k})\\
&u_{l|k}^{*}= \underset{u_{l|k}\in \mathbb{U}}{arg\hspace{.1cm}min} \hspace{.2cm} [L(x_{l|k},u_{l|k})+ V_{l+1|k}^{*}(x_{l+1|k})]
\end{aligned}
	\end{equation}
	where $l=k+N-1,...,k+1,k.$ In MPC, during each time instant only the first element of $U_{k}$ is applied to the system. Consequently, we only need to compute $u_{k|k}$ which can be computed as
	\begin{equation}
	    u_{k|k}^{*}= \underset{u_{k|k}\in \mathbb{U}}{arg\hspace{.1cm}min} \hspace{.2cm} \big[L(x_{k|k},u_{k|k})+ V_{k+1|k}^{*}(x_{k+1|k})\big]
	\end{equation}
This implies that to compute $u_{k|k}$ we need to know $V_{k+1|k}^{*},$ which is the value function from the next instant. To simplify the notations, we denote $u_{k|k}=u_{k}, x_{k|k}=x_{k},x_{k+1|k}=x_{k+1},V_{k|k}=V_{k}, V_{k+1|k}=V_{k+1}$ using which (11) can be rewritten as	
	\begin{equation}
	    u_{k}^{*}= \underset{u_{k}\in \mathbb{U}}{arg\hspace{.1cm}min} \hspace{.2cm} \big[L(x_{k},u_{k})+ V_{k+1}^{*}(x_{k+1})\big]
	\end{equation}	
	
For linear systems with quadratic cost, the value function will be a quadratic function and can be characterized using the Riccati matrix, i.e., $V_{k}=x_{k}^{T}P_{k}x_{k}$ where $P_{k}$ is the Riccati matrix \cite{dp:05}. 	
But for nonlinear systems, there is no general expression for the value function and it is difficult to find an analytical expression for the value function even for the simplest cases.
Therefore we have to go for numerical computation in which we discretize the state and control spaces and compute the value function for each state-control pair. In this case, the computation and storage required for the value function increase exponentially with the length of the time horizon. In general, if there are $n_{x},n_{u}$ possible values for the states and control input (after discretization of the state space and control input space), then the computation cost for the exact value function at time instant $k$ is of order $(n_{x}n_{u})^{N-k}$ (this is known as the curse of dimensionality) \cite{dp:05}.

\subsection{Approximate Dynamic Programming (ADP)} 
One possible solution to the computational complexity of dynamic programming or the curse of dimensionality is the ADP \cite{dp:12}.
Approximate dynamic programming uses an approximation to the value function denoted by $\hat{V}_{k}$ which is usually computed offline. Then the control input at each time instant is computed by solving the approximate Bellman equation
\begin{equation}
\hat{u}_{k}= \underset{u_{k}\in \mathbb{U}}{arg\hspace{.1cm}min} \hspace{.2cm} [L(x_{k},u_{k})+ \hat{V}_{k+1}(x_{k+1})]
\end{equation} 
Note that here the approximate value function $\hat{V}_{k}$ is computed offline, therefore the online computation consists of solving the approximate Bellman equation. If we have $n_{u}$ possible values for the control input, then solving (13) consists of comparing $n_{u}$ values of the cost-to-go function which can be performed easily.
\par In the proposed approach, we use a quadratic approximation to the value function as 
\begin{equation}
    \hat{V}_{k+1}(x_{k+1})=\bar{x}_{k+1}^{T} \bar{P}_{k+1}\bar{x}_{k+1} 
\end{equation}
where $\bar{P}_{k+1}$
 is the parametric matrix which is to be determined. % and $\bar{x}_{k+1}=\left[\begin{matrix}x_{k+1}\\1 \end{matrix}\right]$ is the augmented state vector which will be defined later. 

\section{ADP based MPC Scheme}
%\subsection{Basic Idea}
We denote $\mathbb{P}_{a}$ as the set of matrices used to approximate the value function:
\begin{equation}
  \mathbb{P}_{a}=\{\bar{P}_{1},\bar{P}_{2},...,\bar{P}_{\mu}  \}  
\end{equation}
where $\mu=|\mathbb{P}_{a}|$ and we define $\mathbb{H}=\{1,2,...,\mu\}.$ Then the approximate value function is computed as
\begin{equation}
    \hat{V}_{k+1}(x_{k+1})=\underset{i\in \mathbb{H}}{min} \hspace{.2cm}
    \bar{x}_{k+1}^{T} \bar{P}_{i}\bar{x}_{k+1} \hspace{.5cm}\bar{P}_{i} \in  \mathbb{P}_{a}
\end{equation}
Now the main task is to compute the set $\mathbb{P}_{a}.$ In the proposed approach, we compute $\mathbb{P}_{a}$ offline using the switched system model of the nonlinear system.
%%%%%%%%%%%%%%%%%%%%%%%%%%%%%%%%%%%%%%%%
\subsection{Offline Computation}
In the offline computation part,  the switched system model (7) of the nonlinear system (1) is used to compute the set $\mathbb{P}_{a}.$ 
For a given switching sequence $\sigma=\{\sigma_{0},\sigma_{1},...,\sigma_{N-1}\},$ the switched system (7) becomes a deterministic linear time-varying (LTV) system. For LTV system with a quadratic cost function, the following Lemma can be easily obtained from the standard LQR approach \cite{dp:05}.
\begin{lemma}
The cost-to-go function for the  linear time-varying system  with quadratic cost will be quadratic:
\begin{equation}
    V_{k}=\bar{x}_{k}^{T}\bar{P}_{k}\bar{x}_{k}
\end{equation}
where $\bar{P}_{k}=\bar{Q}_{k}+\bar{A}_{k}^{T}\bar{P}_{k+1} \bar{A}_{k}, \hspace{.3cm} \bar{P}_{N}=\bar{Q}_{N}.$
\end{lemma}
\begin{figure}
	
	\begin{center}
		\includegraphics [scale=.5] {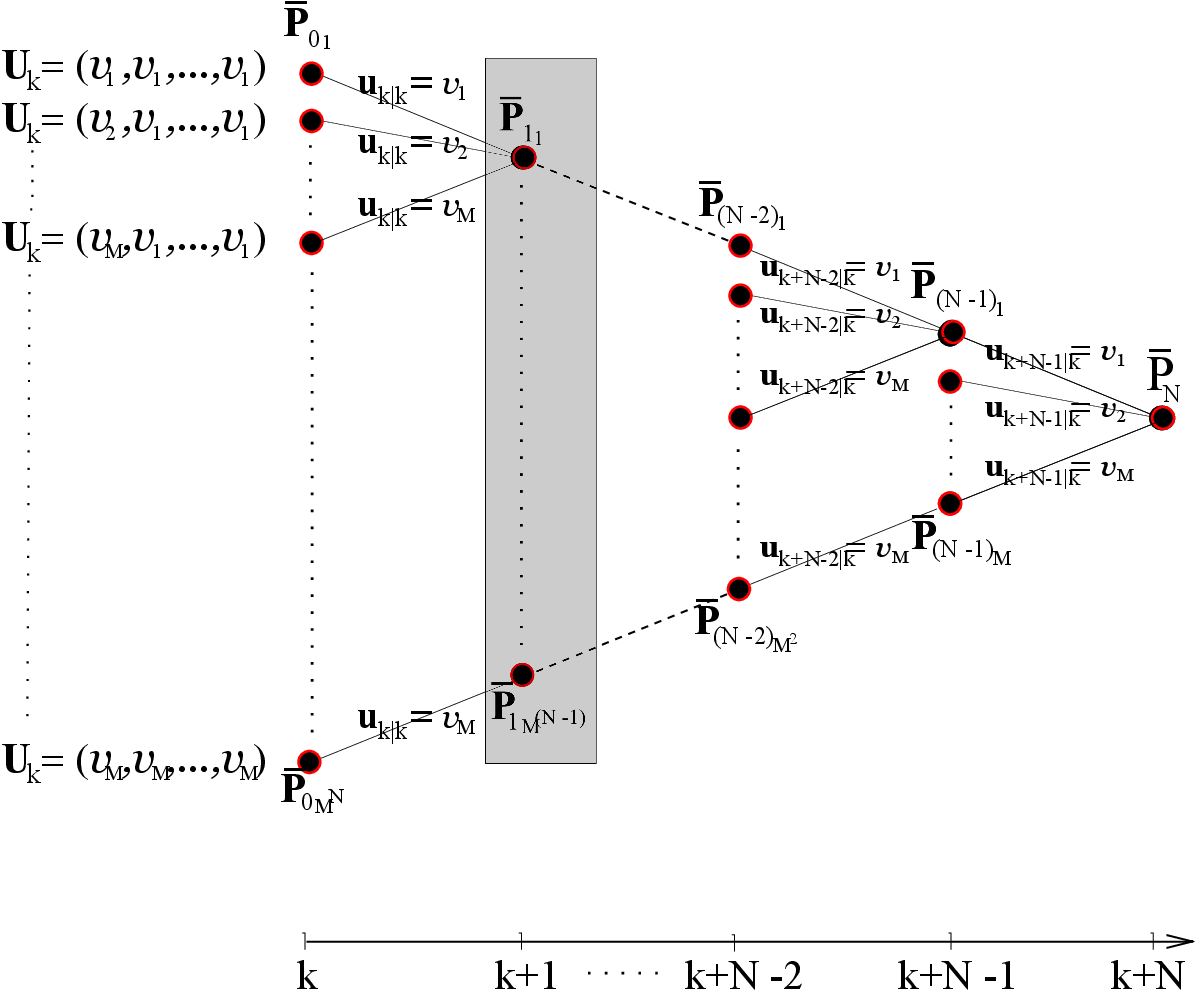}
		\caption{{Switching tree: MPC with prediction horizon $N$}}
		\label{x3}		
	\end{center}
\end{figure}
\par Now for each control sequence $U_{k}\in \mathbb{U}_{v}^{N},$ we can have a switching sequence $\sigma$ and the switched system (7) becomes an LTV system. Since $\mathbb{U}_{v}$ has $M$ number of control inputs, there will be $M^{N}$ control sequences in $\mathbb{U}_{v}^{N}$ which are shown in the switching tree in Fig. 1. For a fixed $N,\bar{P}_{N},$ the set of Riccati matrices for the switching tree will be time invariant. In the proposed MPC scheme, a receding horizon scheme with a fixed $N,\bar{P}_{N}$  is used. 
 Therefore, in  MPC  the value function $V_{1}^{*}$ in the switching tree can be used as an approximation for $V_{k+1}^{*}$ and the value $V_{1}^{*}$ is characterized by the set of Riccati matrices $\mathbb{P}_{1}=\{\bar{P}_{1_{1}},\bar{P}_{1_{2}},...,\bar{P}_{1_{M^{{N}-1}}}\},$ i.e., we have 
\begin{equation}
V_{1}^{*}(x)=\bar{x}^{T} \bar{P}_{1_{i^{*}}}\bar{x}
\end{equation}
where $\bar{P}_{1_{i^{*}}} \in \mathbb{P}_{1}$ and $i^{*}=\underset{i=1,...,M^{N-1}}{arg\hspace{.1cm}min} \hspace{.2cm} \bar{x}^{T}\bar{P}_{1_{i}}\bar{x}.$ In practice, computing and storing $\mathbb{P}_{1}$ will be computationally difficult for problems with larger ${N}.$ Therefore in the proposed method, we use the relaxed pruning scheme introduced \cite{wh:12} for reducing the size of $\mathbb{P}_{1}$ by eliminating the redundant matrices from the set. We denote $\mathbb{P}_{1\epsilon}$ as the $\epsilon-$ equivalent subset of $\mathbb{P}_{1}$ which is obtained by removing the Riccati matrices that are $\epsilon-$ redundant.
We have the following definition for $\epsilon-$ redundancy:

\begin{definition}\cite{wh:12}
For any $\epsilon \geq 0,$ a matrix $\bar{P}_{j} \in \mathbb{P}_{1}$ is called $\epsilon-$ redundant with respect to $\mathbb{P}_{1}$ if, 
\begin{equation}
   \underset{\bar{P} \in  \mathbb{P}_{1} \backslash \bar{P}_{j}}{\mbox{min}} \hspace{.2cm} \bar{x}^{T}\bar{P}\bar{x} \leq \underset{\bar{P} \in  \mathbb{P}_{1}} {\mbox{min}} \hspace{.2cm} \bar{x}^{T}\bar{P}\bar{x} +\epsilon  {\parallel \bar{x} \parallel}^{2}
\end{equation}
for any $\bar{x}\in \mathbb{R}^{n+1}.$
\end{definition}
In that case we prune the matrix ${P}_{j}$ from $\mathbb{P}_{1}$ and the effect of pruning on the value function is the additional cost $\epsilon {\parallel \bar{x} \parallel}^{2}$.
Next lemma \cite{md:21} gives a sufficient condition for $\epsilon-$ redundancy, which is a special case of Lemma 1 in \cite{wh:12}. 
\begin{lemma}
  $\bar{P}_{j} \in \mathbb{P}_{1}$ is $\epsilon-$ redundant if there exists a  $\bar{P}_{i} \in \mathbb{P}_{1}$ that satisfies
\begin{equation}
    \bar{x}^{T}\bar{P}_{i}\bar{x}\leq \bar{x}^{T}\bar{P}_{j}\bar{x}+ \epsilon {\parallel \bar{x} \parallel}^{2} \hspace{.5cm} \forall \bar{x} \in \mathbb{R}^{n+1}
\end{equation}
\end{lemma}

We can rewrite (20) as
\begin{equation}
    \bar{x}^{T}[(\bar{P}_{j}+\epsilon I_{n+1})-\bar{P}_{i}]\bar{x} \geq 0 \hspace{.5cm} \forall \bar{x} \in \mathbb{R}^{n+1}
\end{equation}
This can be verified using any numerical criteria for checking positive definiteness of $(\bar{P}_{j}+\epsilon I_{n+1})-\bar{P}_{i}$ and the  algorithm  for pruning is given in \cite{md:21}.
%{\begin{algorithm}
%	\caption{Offline Final Switching Sequence}
%	\begin{algorithmic}
%		\STATE Input $A_0$, Q, $\Bar{N}$, M, n, $S_1$, $P_1$, $W_1$
%		\STATE c = 0.0001
%		\FOR{i=1:$W_1$}
%		\STATE Calculate E = min $(eig(P_i + cI_n - P_j)) \forall j$
%		\IF {$E \geq 0$ and $j \neq i$}
%	    \STATE $P_1 \leftarrow P_1 \setminus P_i$
%	    \STATE $S_1 \leftarrow S_1 \setminus i$
%	    \STATE $W_0 \leftarrow W_0-1$
%	    \ENDIF
%		\ENDFOR
%		\STATE Output: S_1,P_1,W_1$
%	\end{algorithmic}
%\end{algorithm}

\begin{figure}
\begin{center}
\includegraphics[scale=.25]{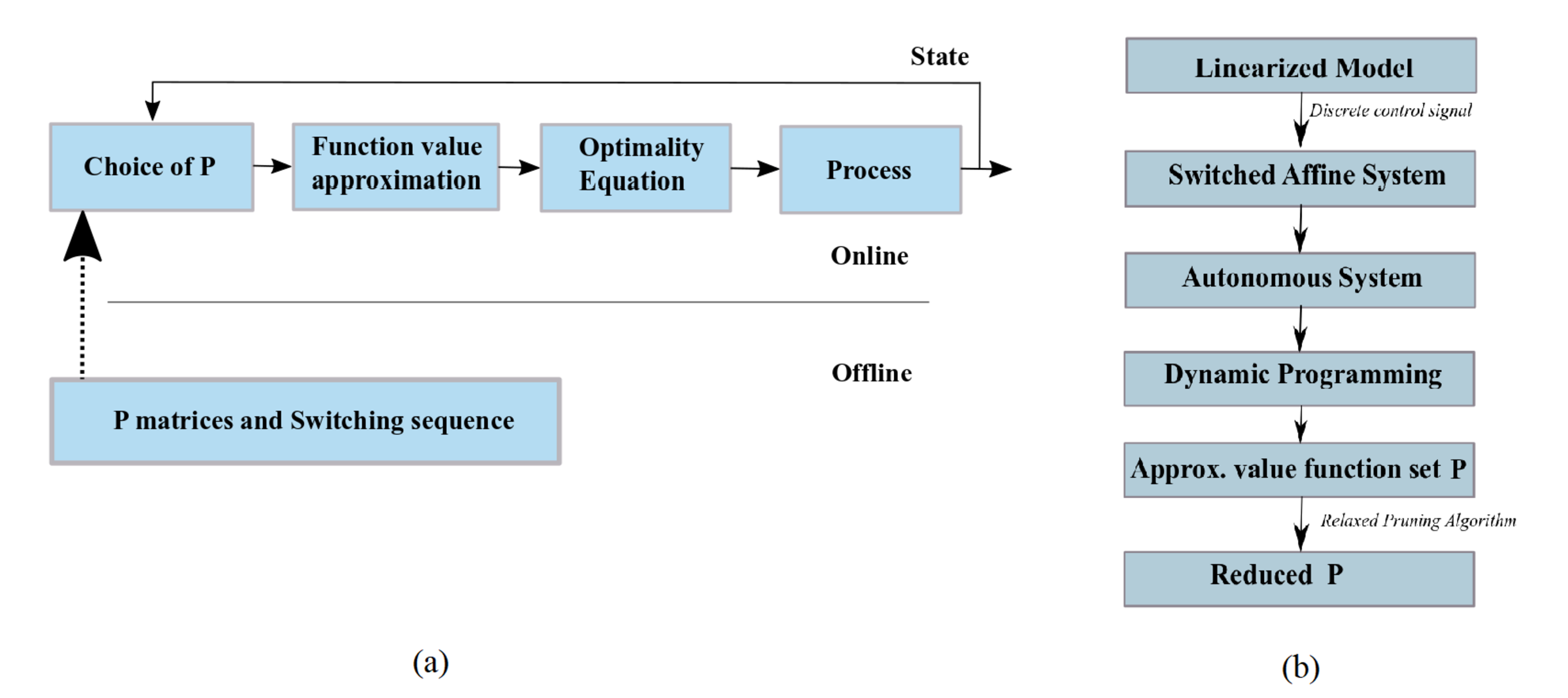}    % The printed column width is 8.4 cm.
\caption{(a) Architecture of the proposed ADP approach (b) Flow chart of the offline Computation} 
\label{figADP}
\end{center}
\end{figure}

In the proposed approach we use $\mathbb{P}_{a}=\mathbb{P}_{1\epsilon}.$ This results in the approximate value function as
\begin{equation}
    \hat{V}_{k+1}(x_{k+1})=\underset{i\in \mathbb{H}}{min} \hspace{.2cm}  \bar{x}_{k+1}^{T} \bar{P}_{i}\bar{x}_{k+1} \hspace{.5cm}\bar{P}_{i} \in  \mathbb{P}_{1\epsilon}
\end{equation}
where $\mathbb{H}=\{1,2,...,\mu\},\hspace{.1cm}\mu=|\mathbb{P}_{1\epsilon}|.$ The architecture of the proposed ADP approach and the flowchart of the offline computation are given in Fig. \ref{figADP}.

%%%%%%%%%%%%%%%%%%%%%%%%%%%%%%%%%%%%%%%%%%%%%
\subsection{Online Computation}
\subsubsection{ADP based MPC Scheme: Single-stage}
 In ADP based MPC scheme, a suboptimal control input is computed at each instant by solving 
		\begin{equation}\label{proposal}
\hat{u}_{k}=v_{k}^{*}=   \underset{v\in \mathbb{U}_{v}}{arg\hspace{.1cm}min} \hspace{.2cm} [L(x_{k},v)+ \hat{V}_{k+1}(x_{k+1})] 
\end{equation} 
	where $\hat{V}_{k+1}$ is the approximate value function computed using (22). Here for each control input $v\in \mathbb{U}_{v},$ we have to find the approximate value function $\hat{V}_{k+1}(x_{k+1}).$ Now $\mathbb{U}_{v}$ contains $M$ number of control inputs and to compute $\hat{V}_{k+1}$ we have to compare $\mu$ number of cost values. Therefore the online computation required for the MPC scheme is of order $\mu M.$
\begin{algorithm}[H]
	\caption{Online computation}
	\begin{algorithmic}
		\STATE Input $x_k$, k, ${N}$
		\FOR  {$j= 1~to~ M $}
		\STATE Calculate $x_{k+1|v_j}$
	    \FOR {$i= 1~to~ \mu $}
	    \STATE $V_{i}= \hat{V}_{k+1}(x_{k+1|v_j})$
	    \ENDFOR
	   \STATE $V^{*}$ = min(V)
	   \STATE $J_{j}=L(x_{k},v_{j})+V^{*}$
	    \ENDFOR
	    \STATE $q$=arg min (J)
	   \STATE $u_{k}=v_{q}$ 
		\STATE Apply $u_k$
		\STATE %Output: $S_1,P_1,W_1$
	\end{algorithmic}
\end{algorithm}

%\begin{figure}
%\begin{center}
%\includegraphics[width=11cm]{flowchart_offline.png}    % The printed %column width is 8.4 cm.
%\caption{Flow Chart of Offline Computations} 
%\label{fig:bifurcation}
%\end{center}
%\end{figure}
%\subsubsection{Issues of Single Stage ADP}

\subsubsection{ADP based MPC Scheme: Multi-stage}
The control law (\ref{proposal}) can be improved by using a multi-stage scheme, in which the control input is computed by considering additional samples around $v_{k}^{*}$ and optimizing over the samples. We denote $\mathbb{U}_{w}=\{w_{-W},...,w_{-1},w_{0},w_{1},...,w_{W}\},$ $ w_{0}=0,$ which consist of $2W+1$ samples around zero. In the case of uniform sampling with a sampling interval $\varDelta u,$ we select $w_{q}=q \varDelta u, q=-W,..-1,0,1,...,W.$ Then in the second stage of optimization, the decision variable becomes $v_{k}^{*}+w$ where $w\in \mathbb{U}_{w}$
and results in the control input:
	\begin{equation}
\hat{u}_{k}=v_{k}^{*}+w_{k}^{*}=v_{k}^{*}+   \underset{w\in \mathbb{U}_{w}}{arg\hspace{.1cm}min} \hspace{.2cm} [L(x_{k},v_{k}^{*}+w)+ \hat{V}_{k+1}(x_{k+1})] 
\end{equation} 
to reduce the online computation we use the same $\hat{V}_{k+1}$ (i.e., same $P$ matrix) obtained in stage 1.

%\begin{algorithm}
%	\caption{Offline Initial Switching Sequence}
%	\begin{algorithmic}
%		\STATE Input $A_0$, Q, $\Bar{N}$, M, n
%		\STATE Set $S_1$, $P_1$, $W_1$ = 0 
%		\FOR{i=$\Bar{N}$:-1: 1}
%	    \STATE $P = Q + A_i^T PA_i$ $ \forall ~i \in M$
%	    \STATE $P_1 \leftarrow P_1 \cup P$
%	    \STATE $S_1 \leftarrow S_1 \cup i$
%	    \STATE $W_1 \leftarrow W_1+1$
%		\ENDFOR
%		\STATE Output: $S_1,P_1,W_1$
%	\end{algorithmic}
%\end{algorithm}

%%%%%%%%%%%%%%%%%%%%%%%%%%%%%%%% CONVERGENCE ANALYSIS %%%%%%%%%%%%%%%%%%%%
\section{MPC with bounded state constraints}
So far we considered the MPC problem with an unconstrained state case, i.e., only the control input is constrained. In this section, we extend the ADP-based MPC scheme to nonlinear systems with both input and state constraints (bounded). We consider ${x}_{k}\in \mathbb{X},$ ${u}_{k}\in \mathbb{U},$ and $\mathbb{X},$ $\mathbb{U}$ are considered as polytopes in $\mathbb{R}^{n}, \mathbb{R}^{m}$ respectively. 
\par In the unconstrained state case we used the set of matrices $\mathbb{P}_{a}$ to approximate the value function. The set $\mathbb{P}_{a}$ is computed by pruning the set of initial matrices of the switching tree, and the pruning criteria is needed to be satisfied for all values of the state $\bar{x}.$ Now in the constrained state case we can use the set $\mathbb{P}_{a}$ and the size of $\mathbb{P}_{a}$ can be reduced further by considering the matrices that result in an optimal cost function for the states in the region $\mathbb{X},$ instead of the entire state space $\mathbb{R}^{n}.$ 
We define the set $\mathbb{P}_{a_d}$ as the set of initial matrices for the constraint state case.
Then to compute $\mathbb{P}_{a_d},$ we discretize the constraint set $\mathbb{X}$ into $n_{\mathbb{X}}$ points and the set is denoted as $\mathbb{X}_{d}=\{x_{d_1},x_{d_2},...,x_{d_{n_{\mathbb{X}}}}\}\subset \mathbb{X}.$ Then for each ${x}_{d}\in \mathbb{X}_{d}$ we use the augmented state $\bar{x}_{d}$ to find the optimal Riccati matrix from $\mathbb{P}_{a}$ as below
\begin{equation}
        \bar{P}^{*}=\underset{\bar{P}_{i}\in \mathbb{P}_{a}}{arg\hspace{.1cm}min} \hspace{.2cm}  \bar{x}_{d}^{T} \bar{P}_{i}\bar{x}_{d} \hspace{.5cm}\bar{P}_{i} \in  \mathbb{P}_{a}, i\in \{1,2,...,\mu\}
\end{equation}
and each time when we get a new $\bar{P}^{*}$ we add it to the set $\mathbb{P}_{a_d}.$ Clearly $\mathbb{P}_{a_d} \subseteq \mathbb{P}_{a}$ and in the constraint state case we use the set $\mathbb{P}_{a_d}$ for the computation of $\hat{V}_{k+1}$.
 Then the control input for the MPC is computed at each instant by solving 
		\begin{equation}
		\begin{aligned}
&\hat{u}_{k}=v_{k}^{*}=   \underset{v\in \mathbb{U}_{v}}{arg\hspace{.1cm}min} \hspace{.2cm} [L(x_{k},v)+ \hat{V}_{k+1}(x_{k+1})]\\
&subject\hspace{.1cm} to\hspace{.2cm} f(x_{k},v)\in \mathbb{X}
\end{aligned}
\end{equation} 
%where $\hat{V}_{k+1}$ is the approximate value function computed using (22) with $\mathbb{P}_{a}=\mathbb{P}_{a_d}$.
\par The online computation can be reduced further by dividing the constraint set $\mathbb{X}$ into a finite number of polytopes $\mathbb{X}_{1},\mathbb{X}_{2},...\mathbb{X}_{z}$ and then find the set of initial matrices for each of these regions, which results in $\mathbb{P}_{a_{d_1}},\mathbb{P}_{a_{d_2}},...,\mathbb{P}_{a_{d_{z}}}.$ Then during the online computation, if $x_{k}\in \mathbb{X}_{j}$ we use the set $\mathbb{P}_{a_{d_{j}}}$ for approximating the value function. Here we have $\mathbb{P}_{a_{d_j}} \subseteq \mathbb{P}_{a_d}$ for each $j\in \{1,2,...,z\}.$ Therefore this approach can be used for reducing online computation.

\section{Stability Analysis}
Next, we discuss the stability of the proposed MPC scheme in which we focus on a posteriori stability analysis. In general, stability of MPC schemes is achieved by considering the optimal cost function $J_{N}^{*}(x_{k})$ as a candidate Lyapunov function. 
For the proposed MPC scheme (\ref{proposal}) we obtained the cost $J_{N}^{*}(x_{k})$ as
\begin{equation}
    \begin{aligned}
    J_{N}^{*}(x_{k})&=\underset{v\in \mathbb{U}_{v}}{min} \hspace{.2cm} [L(x_{k},v)+ \hat{V}_{k+1}(x_{k+1})] =L(x_{k},v^{*})+  \bar{x}_{k+1}^{T} \bar{P}_{{i}^{*}}\bar{x}_{k+1}\\
       &=L(x_{k},v^{*})+\left[\begin{matrix}
f(x_{k},v^{*})\\ 1
\end{matrix}\right]^{T} \bar{P}_{{i}^{*}}\left[\begin{matrix}
f(x_{k},v^{*})\\ 1
\end{matrix}\right]
    \end{aligned}
\end{equation}
which is a nonlinear function of $x_k$ and can be non-quadratic as well  because of the nonlinear mapping $f(x_k,v^{*}).$ This complicates the stability analysis of the proposed MPC scheme since proving $J_{N}^{*}$ will be a Lyapunov function for the nonlinear system a priori will be difficult. This motivates us to go for a posteriori stability analysis \cite{tg:08}, in which we give sufficient conditions for  $J_{N}^{*}$ to be a Lyapunov function for the nonlinear system which can be verified numerically. 
\par In the posteriori analysis, we aim to numerically show that for the proposed MPC scheme the cost $J_{N}^{*}$ will be a Lyapunov function so that the corresponding closed-loop dynamics is asymptotically stable. %We have considered extended state to facilitate the tracking control problem. 
Before proceeding further, the following assumption is made:
%%%% Before proceeding further, the following assumption is made: %%%%
\begin{assumption}
The nonlinear function $f(x,u)$ is continuous and locally Lypchitz with respect to $x$ in $\mathbb{X}\times \mathbb{U},$ i.e., there exists a constant $L_{c}\in \mathbb{R}_{+}$ such that
\begin{equation}
\parallel    f(x_{1},u)-f(x_{2},u) \parallel \leq L_{c} \parallel x_{1}-x_{2} \parallel, \hspace{.5cm} \forall x_{1},x_{2} \in \mathbb{X}, u \in \mathbb{U}.
\end{equation}
\end{assumption}
Assumption 1 ensures the mapping $f(x,u)$ does not go unbounded for small changes of $x.$ %%%%%% Furthermore, the stage cost function $L(x_{k},v^{*})$ is local Lipschitz with respect to x,u in the domain of $\mathbb{X}\times \mathbb{U}$. %%%%%
Next, we analyze the stability of the nonlinear system and we had transformed the system in such a way that the origin ${x}_k=0$ will be the equilibrium point. We have the following theorem on stability \cite{tg:08}
\begin{theorem} 
The nonlinear system (1) with the MPC scheme is asymptotically stable on $\mathbb{X}$, if there exists a Lyapunov function $L_{f}(x_k)$ which satisfies 
%\begin{equation}
%    L(\bar{x}_k)=\bar{x}^T_k P_i \bar{x}_k ~~if ~~\bar{x}_k \in \mathbb{X}
%\end{equation}
%where $P_i$ is positive definite matrix from the set $\mathbb{P}$ and \mathbb{X} is a polyhedron.

\begin{equation}
    L_{f}({x}_k) \geq c_{1} ||{x}_k||^2_2, \hspace{.5cm} \forall \hspace{.1cm} {x}_k \in \mathbb{X}\backslash \{0\}, \hspace{.2cm} c_1 > 0, \hspace{.2cm}  L_{f}(0)=0
\end{equation}
and 
\begin{equation}
     L_{f}({x}_{k+1}) -  L_{f}({x}_k) \leq - c_2 ||{x}_k||^2_2, \hspace{.5cm} \forall \hspace{.1cm} {x}_k \in \mathbb{X}\backslash \{0\}, \hspace{.2cm} c_2 >0.
\end{equation}
\end{theorem}
For the proposed MPC scheme, the value function is considered as a candidate Lyapunov function, i.e. $ L_{f}(x_k)=J_{N}^{*}(x_k).$
If the above conditions are satisfied for $ L_{f}(x)=J_{N}^{*}(x),$ then the nonlinear system with the MPC scheme is asymptotically stable within the constrained set $\mathbb{X}$. Clearly, the first condition (29) is satisfied by $J_{N}^{*}(x_k)$ with $c_1=\lambda_{min}(Q),$ since $\bar{P}_{i^*} \geq 0$ in (27).
In order to verify the second condition,  we discretize $\mathbb{X}$ into $\mathbb{X}_{d}\subset \mathbb{X}\backslash \{0\}$ and for each $x\in \mathbb{X}_{d}$ check whether $J_{N}^{*}$ satisfies the condition (30). By considering sufficiently large number of points in $\mathbb{X}_{d},$ we can find $c_2$  with a desired accuracy. For example if we discretize each component of the state vector with a step size $10^{-h}$ then the posteriori stability bounds computed using $\mathbb{X}_{d}$  holds up to $h$ digit precision. In order to compute $c_2,$ for each $x\in \mathbb{X}_{d},$ we compute
\begin{equation}
 c(x)=\frac{    L_{f}\big(f(x,v^{*})\big) -  L_{f}({x}) }{ ||{x}||^2_2}= \frac{   J_{N}^{*}\big(f(x,v^{*})\big) - J_{N}^{*}({x}) }{ ||{x}||^2_2}
\end{equation}
from which we obtain
\begin{equation}
    -c_{2}=\underset{x\in \mathbb{X}_{d}}{max} \hspace{.2cm}c(x).
\end{equation}
Now, if $-c_{2}<0$ (i.e., $c_{2}>0$), then $J_{N}^{*}$ satisfies (30) and can be used as a Lyapunov function for the nonlinear system with the MPC scheme.
%%%%%%%%%%%%%%%%%%%%%%%%%%%%%%%%%%%% Multi-Tank System %%%%%%%%%%%%%%%%%%%

\section{Experimental study:  Multi-Tank System}

\begin{figure}
\begin{center}
\includegraphics[width=16cm]{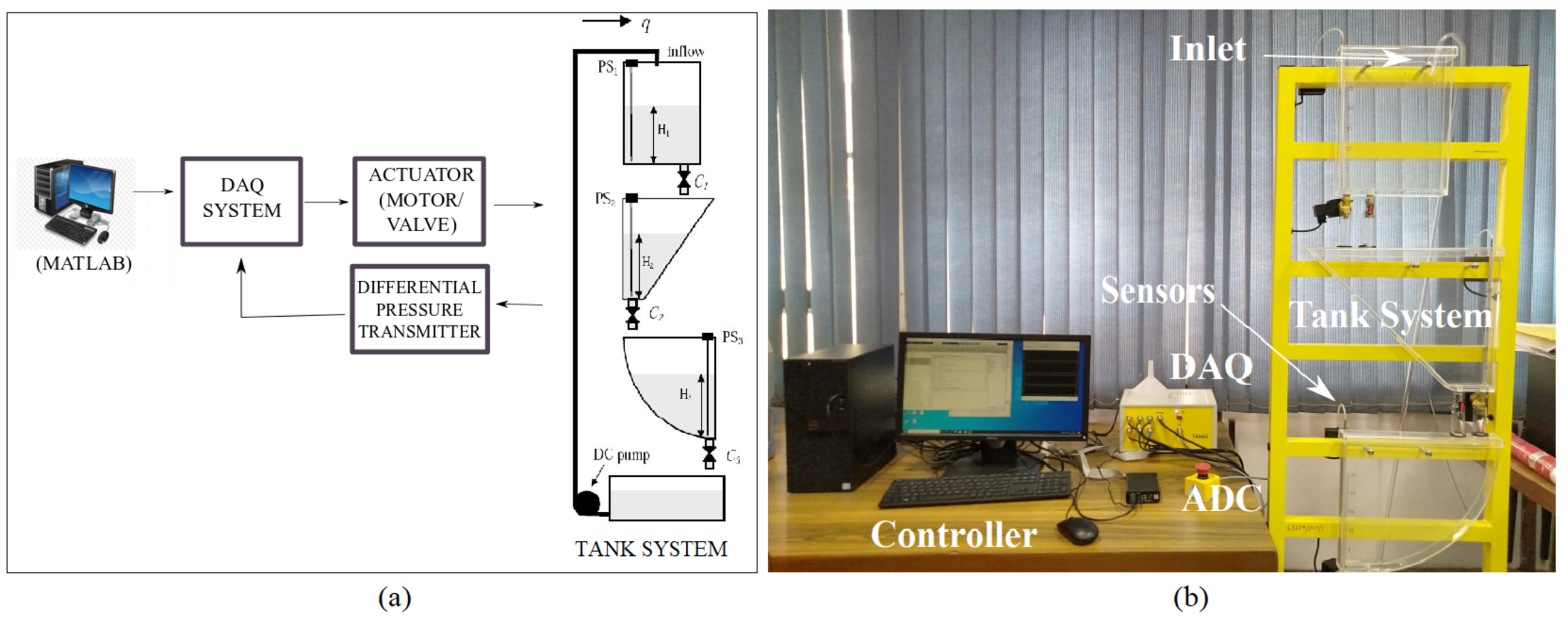}    % The printed column width is 8.4 cm.
\caption{Multi-Tank Control System (a) Block diagram (b) Experimental setup} 
\label{fig:block}
\end{center}
\end{figure}

Figure \ref{fig:block}(a) shows the block diagram of the multi-tank system and \ref{fig:block}(b) shows the actual experimental setup. Our target is to maintain the water levels of the three tanks at the desired values. In the computer, we specify the set point and load the control algorithm. The water levels $H_1$,$H_2$, and $H_3$ are controlled either by controlling the valve or by controlling the speed of the motor. In this paper, we keep the valve position fixed and give the control signal to vary the speed of the motor that pumps water to the tank. The differential pressure sensor measures the water level and sends it back to the computer through the data acquisition system.

In the multi-tank system, we pump the water to the top tank from the reservoir located at the bottom. The liquid inflow $q$ is controlled by the pump. The valve $C_1$ restricts the flow of the liquid from the topmost tank to the intermediate tank. The water flows down due to gravity from the second tank to the third tank passing through valve $C_2$. And the liquid finally reaches the reservoir through valve $C_3$. The computer communicates with the level sensors, pumps, and valves by a dedicated input-output board and power interface \cite{ind:03}. A real-time software operating in MATLAB Real-Time Windows Target (RTWT) rapid prototyping environment controls the input-output board. The MATLAB version on the computer is R2018b and  Intel Core i7 processor at 3GHz with 32GB RAM. The run time of the experiment is 250 sec with a sampling time of 0.01 sec.

 %Figure \ref{fig:block}(a) shows the block diagram of the multi-tank experimental setup. The set-point is given in the computer, where the control algorithm is loaded. The control signal is sent to the actuator, which can be the valve or motor, to control the tank's level. The water level is measured by a differential pressure transmitter and sent back to the computer through the data acquisition system.
 
%Figure \ref{fig:block} shows the schematic diagram of the Multi-Tank system.  The bottom part is the reservoir from where water is pumped to the top tank. The DC pump controls the liquid inflow $q$. The water from the top tank then flows down through the valve $C_1$ to tank 2 due to gravity, and from there, water flows down to tank 3 through valve $C_2$. This water then reaches the reservoir through valve $C_3$. The valves resist the flow. We aim to control the three tanks' water levels $H_1$,$H_2$,$H_3$. The Computer can give the control input to valves or the motor. We keep the valves at fixed positions, and the control input adjusts the motor's speed. This variable-speed motor fills the upper tank.

%\begin{figure}
%\begin{center}
%\includegraphics[width=4.4cm]{Multi_Tank_Configuration.JPG}    % The printed column width is 8.4 cm.
%\caption{Schematic Diagram of Multi Tank System} 
%\label{fig:schematic}
%\end{center}
%\end{figure}

%Figure 3 shows the motor control of the three-tank system. 

%\begin{figure}
%\begin{center}
%\includegraphics[width=11cm]{motor_control.JPG}    % The printed column width is 8.4 cm.
%\caption{Block diagram of Motor Control of Three Tank} 
%\label{fig:bifurcation}
%\end{center}
%\end{figure}
%%%%%%%%%%%%%%%%%%%%%%%%%%%%%%%%%%%%%%
\subsection{System Equations}

Bernoulli's law governs the outflow rate of the fluid from a tank. The outflow rate of a tank is given by:
\begin{equation}
Q_r = f_0 S \sqrt{2gH_0}
\end{equation} 
where $f_0$ is orifice outflow coefficient and S is the output area of orifice.\\
 The nonlinear model of the multi-tank system \cite{ind:03} is given by:
 
 %\begin{equation}
 %\begin{split}
 \begin{align}
    \frac{dH_1}{dt} &= \frac{1}{\beta_1(H_1)}q - \frac{1}{\beta_1(H_1)}C_1 H_1^{\alpha_1}\\
 \frac{dH_2}{dt} &= \frac{1}{\beta_2(H_2)}C_1 H_1^{\alpha_1} - \frac{1}{\beta_2(H_2)}C_2 H_2^{\alpha_2}\\
 \frac{dH_3}{dt} &= \frac{1}{\beta_3(H_3)}C_2 H_2^{\alpha_2} - \frac{1}{\beta_3(H_3)}C_3 H_3^{\alpha_3}
 \end{align}
%\end{split}
% \end{equation}
where height of fluid, resistance of output orifice, flow coefficients are $H_i$, $C_i$, $\alpha_i$ respectively  for $i-{th}$ tank, $i=1,2,3$. The parameter values are given in the next section.\\
$\beta_1(H_1) = aw_t $ - Cross sectional area of topmost tank; 
$\beta_2(H_2) = cw_t + \frac{H_2}{H_{2max}}bw_t $ - Cross sectional area of intermediate tank; 
$\beta_3(H_3) = w_t\sqrt{R^2-(R-H_3)^2} $ - Cross sectional area of bottom tank.
The variable cross-section of the tanks, the contribution of valve geometry \& flow dynamics, and the pump \& valve input-output characteristics are the potential factors that challenge the control design.

%%%%%%%%%%%%%%%%%%%%%%%%%%%%%%%%%%%%%%%%%%%%%%

\subsection{Model and Parameters}
The linearized dynamical model of the three-tank system, linearized around ($H_{10}, H_{20}, H_{30}$), is described as: $h_{k+1}=Ah_k+Bu_k$

\begin{equation}
A=\begin{bmatrix}
\frac{-C_1 \alpha_1}{aw_tH_{10}^{1-\alpha_1}} & 0 & 0 \\
\frac{C_1 \alpha_1}{w_t\bigg(c+b\frac{H_20}{H_{2max}}\bigg) H_{10}^{1-\alpha_1}} &  \frac{-C_2 \alpha_2}{w_t\bigg(c+b\frac{H_20}{H_{2max}}\bigg) H_{20}^{1-\alpha_2}} & 0 \\
0 &\frac{C_2 \alpha_2}{w_t \sqrt{R^2 - (H_{3max}-H_30)^2} H_{20}^{1-\alpha_2}}  &\frac{-C_3 \alpha_3}{w_t \sqrt{R^2 - (H_{3max}-H_30)^2} H_{30}^{1-\alpha_3}}
\end{bmatrix}, \hspace{.1cm}
B=\begin{bmatrix}
\frac{1}{aw_t} \\
0\\
0
\end{bmatrix}
\end{equation}

The experimental setup uses the parameter values $H_{1max},H_{2max},H_{3max}$ = 0.35 m, $a$ = 0.25 m, $b$ = 0.345 m, $w_t$ = 0.035 m, $c$ = 0.10 m, $R$ = 0.365 m. The constraints are the maximum height of the tank is 0.35 m and the control input could be varied from 0.54 to 1 volt corresponding to the minimum and the maximum motor speed respectively.We discrete (quantized) set of control inputs into $\mathbb{U}_{v} =\{v_{1},v_{2},...,v_{6}\}$ and $\mathbb{U}_{v}\subset \mathbb{U}.$ The LTI system with discrete set of controls $\mathbb{U}_{v}$ is represented as a switched affine system as in (6) with $M=6$ subsystems.

%%%%%%%%%%%% RESULT AND COMPARISON %%%%%%%%%%%%%%%%%%%%%%%%%%

\subsection{Results}
\begin{figure}
\begin{center}
\includegraphics[width=16cm]{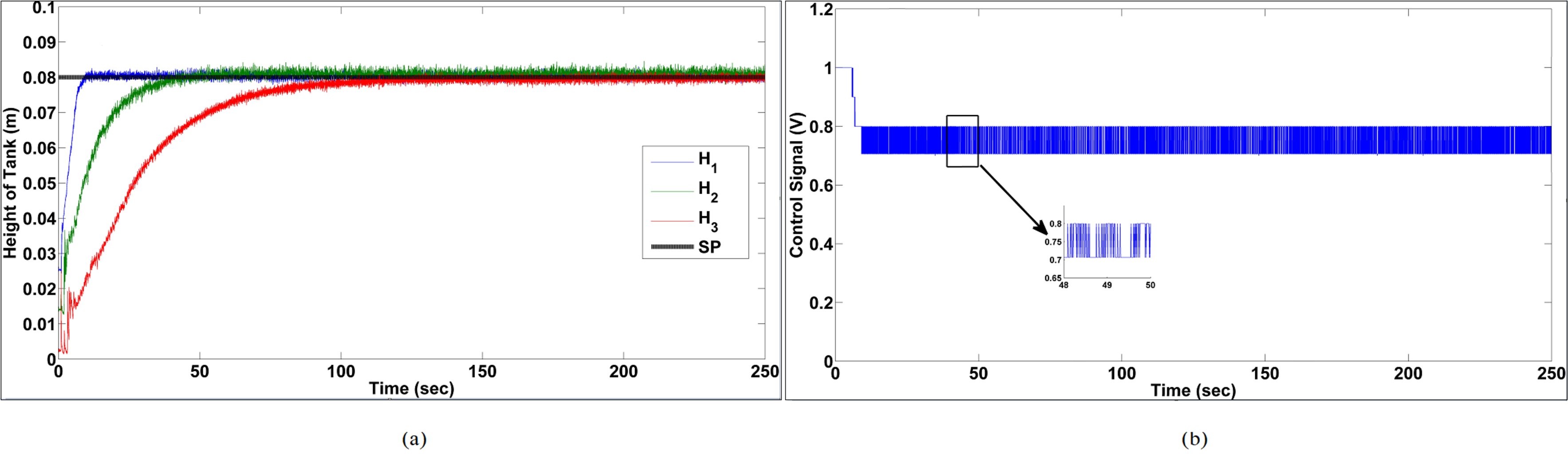}    % The printed column width is 8.4 cm.
\caption{Single-stage ADP (a) Evolution of states (b) Control signal} 
\label{STATES_SINGLESTAGE}
\end{center}
\end{figure}

We have implemented algorithm 1 in the setup. Figure \ref{STATES_SINGLESTAGE}(a) shows the responses of single-stage ADP (ADP-1) when applied to the multi-tank system. The corresponding control signal is shown in figure \ref{STATES_SINGLESTAGE}(b). We observe continuous chattering of the control signal due to the disturbance in state measurement. 

%\begin{figure}
%\begin{center}
%\includegraphics[width=11cm]{REAL_SETUP.pdf}    % The printed column width is 8.4 cm.
%\caption{Experimental Setup: Multi Tank System} 
%\label{STATES_SINGLESTAGE}
%\end{center}
%\end{figure}

\begin{figure}
\begin{center}
\includegraphics[width=16cm]{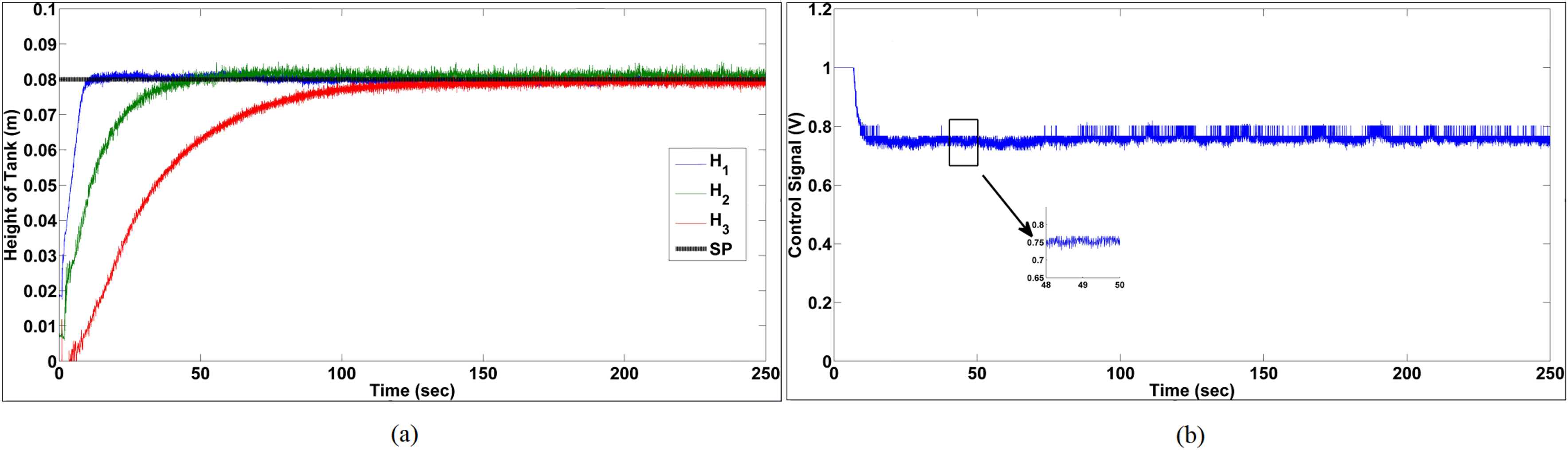}    % The printed column width is 8.4 cm.
\caption{Multi-stage ADP (a) Evolution of states (b) Control signal} 
\label{states_multistage}
\end{center}
\end{figure}

A more accurate control at each instant is obtained by the use of multi-stage ADP (ADP-2). The evolution of states for multi-stage ADP applied to the multi-tank system is shown in figure  \ref{states_multistage}(a). And the corresponding control signal is shown in figure \ref{states_multistage}(b). We notice that the control signal in figure \ref{states_multistage}(b) has become more smooth compared to figure \ref{STATES_SINGLESTAGE}(b).

%\begin{itemize}
% \item Comparison of Average computation time - Online and Offline
%\item Comparison of Average computation time - Online and Offline
%\item Performance comparison - ISE, IAE, ITAE
%\end{itemize}

%\begin{figure}
%\begin{center}
%\includegraphics[width=11cm]{SINGLESTAGE_CONTROL_30J.jpg}    % The printed column width is 8.4 cm.
%\caption{Control Signal: Single Stage ADP} 
%\label{CONTROL_SINGLESTAGE}
%\end{center}
%\end{figure}

\begin{figure}
\begin{center}
\includegraphics[width=10cm]{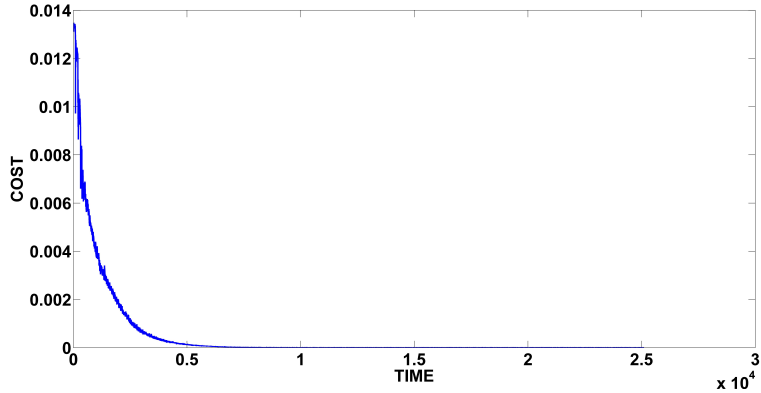}    % The printed column width is 8.4 cm.
\caption{Value of Cost function along closed loop trajectory} 
\label{value_fcn}
\end{center}
\end{figure}

%\begin{figure}
%\begin{center}
%\includegraphics[width=11cm]{MULTISTAGE_CONTROL_30J.jpg}    % The printed column width is 8.4 cm.
%\caption{Control Signal: Multi Stage ADP} 
%\label{control_multistage}
%\end{center}
%\end{figure}

%\subsection{Analysis}
By carrying out a posteriori analysis, we observed that the proposed controller leads to asymptotical stability. %The equilibrium point of the system is asymptotically stable on $\mathbb{X}_0$ if there exists a Lyapunov function as a cost function.
For the proposed MPC, we have obtained $c_{2}=0.4318$ for the Lyapunov function $L_{f}(x_{k})=J_{N}^{*}(x_{k}).$ %for all initial states within the state constraints.
This is obtained by discretizing the constraint set $\mathbb{X}$  in to $\mathbb{X}_{d}$ which contains $10^9$ points (step size $10^{-3}$, i.e. $h=3$) and computing $c(x)$ as in (31) for each $x\in \mathbb{X}_{d}.$ From which the maximum value gives $c_{2}$ and for the multi-tank system with proposed MPC scheme we obtained $c_{2}=0.4318>0$. Consequently, the cost $J_{N}^{*}(x_{k})$ becomes a Lyapunov function and the proposed MPC scheme results in asymptotically stable closed loop dynamics with $3$ digit precision as per Theorem 1. From Fig. \ref{value_fcn}, we can see that the value function is decreasing along the trajectory. We observe that the performance function reduces to zero as the states reach the equilibrium point.
\subsection{Comparison}
The comparison of the computation time is shown in table \ref{tb:comparison}. The computation of Nonlinear Model Predictive Control (NMPC) is more (27\%) compared to ADP-based MPC. ADP-1 is the single-stage Approximated Dynamic Programming based MPC and ADP-2 is the multi-stage ADP based MPC. The computation time of ADP-2 is slightly higher (0.39\%) compared to ADP-1 due to the additional computation time taken for the second stage of iteration to find a more suitable control signal. But in multi-stage ADP we get smoother control compared to a single-stage as observed from \ref{STATES_SINGLESTAGE}(b) and \ref{states_multistage}(b). The ADP-3 shows the case where we considered the state constraints. The computation time got further reduced as we were able to reduce the set of P matrices. The table \ref{tb:comparison} also shows the comparison of Integral Square Error (ISE=${\sum_{k=0}^T} {||e_{k}||}_{2}^2$; where $e_{k}=x_k-x_r$ ) for different strategies. The value is least for NMPC, but the proposed strategies have comparable performance compared to NMPC.
\begin{center}
\begin{table}%
\centering
\caption{Comparison of Different Strategies\label{tb:comparison}}%
\begin{tabular*}{500pt}{@{\extracolsep\fill}lccc@{\extracolsep\fill}}
\toprule
\textbf{Control Scheme} & \textbf{Time (sec)}  & \textbf{ISE}  \\
\midrule
NMPC & 0.0012 & 10.3225\\
ADP-1 & $4.2810 \times 10^{-5}$ & 12.3790\\
ADP-2 & $7.0993 \times 10^{-5}$ & 11.5444\\
ADP-3 & $9.1070 \times 10^{-6}$ & 13.2071\\
\bottomrule
\end{tabular*}
\end{table}
\end{center}
%\begin{table}[hb]
%\begin{center}
%\caption{Comparison of Different Strategies}\label{tb:comparison}
%\begin{tabular}{|c|c|c|}\hline
%Control Scheme & Time (sec) & ISE\\\hline
%NMPC & 0.0012 & 10.3225\\
%ADP-1 & $4.2810 \times 10^{-5}$ & 12.3790\\
%ADP-2 & $7.0993 \times 10^{-5}$ & 11.5444\\
%ADP-3 & $9.1070 \times 10^{-6}$ & 13.2071\\ \hline
%%Offset  \\
%%ISE/IAE/ITAE \\
%\end{tabular}
%\end{center}
%\end{table}

\section{Conclusion}
In this paper, we have proposed an approximate dynamic programming-based MPC to reduce the online computation time. To evaluate the advantage of the proposed schemes they were applied to a multi-tank nonlinear system. A smoother control was obtained in the multi-stage MPC scheme compared to the single-stage MPC scheme. The computation time was further reduced by considering state bounds. The method reduces the computation time significantly, while the performance in terms of ISE is comparable.

%%%%%%%%%%%%%%%%%%%%%%%%%%%% THEOREM %%%%%%%%%%%%%%%%%%%%%%%%%%%%%%%

%\begin{theorem}[Theorem subhead]\label{thm1}

%\end{theorem}

%%%%%%%%%%%%%%%%%%%%%%%%%%%%%%%%% definition %%%%%%%%%%%%%%%%%%%%%%%
%\begin{definition}[Definition sub head]
%
%\end{definition}

%\backmatter

%\section*{Acknowledgments}

\bibliographystyle{unsrt}

\begin{thebibliography}{1}

     \bibitem{kb:15}
   K. Basil, C. Mark, \emph{\textquotedblleft Model Predictive Control- Classical, Robust and Stochastic\textquotedblright,} Springer, 2015. 

     \bibitem{jml:05}
   J. M. Lee and J. H. Lee, \emph{\textquotedblleft Approximate dynamic programming-based approaches for input–output data-driven control of nonlinear processes\textquotedblright,} Automatica, vol. 41, pp. 1281-1288, 2005.

 

  \bibitem{fl:12}
  F. Lewis and D. Liu, \emph{\textquotedblleft Reinforcement Learning and Approximate Dynamic Programming for Feedback Control\textquotedblright,} John Wiley and Sons,  2012.


   \bibitem{pr:21}
P. Rokhforoz, H Kebriaei, and M. Ahmadabadi,\emph{\textquotedblleft  Large scale dynamic system optimization using dual decomposition method with approximate dynamic programming,\textquotedblright} 
Systems and Control Letters, vol.  150, pp. 550-565,   2021.



 \bibitem{adk:21}
A. Keyser, H. Vansompel, and G. Crevecoeuri,\emph{\textquotedblleft Real-Time Energy-Efficient Actuation of Induction Motor Drives Using Approximate Dynamic Programming,\textquotedblright} 
IEEE Transactions on Industrial Electronics, vol. 68, 2021.





 \bibitem{jl:04}
   J. M. Lee and J. H. Lee, \emph{\textquotedblleft Approximate Dynamic Programming Strategies and Their Applicability for Process Control: A Review and Future Directions\textquotedblright,} International Journal of Control, Automation, and Systems, vol. 3, pp. 263-278, 2004.




     \bibitem{ld:19}
   L. Dong, J. Yan, X. Yuan, H. He, and C. Sun, \emph{\textquotedblleft Functional Nonlinear Model Predictive Control Based on Adaptive Dynamic Programming\textquotedblright,} IEEE Transactions On Cybernetics, vol. 49, pp. 23-35, 2019.

    

     \bibitem{nz:20}
   N. Zhang, B. Leibowicz, and G. Hanasusanto, \emph{\textquotedblleft Optimal Residential Battery Storage Operations Using Robust Data-Driven Dynamic Programming\textquotedblright,} IEEE Transactions on Smart Grid, vol. 11, pp. 1771-1780, 2020.

   
   \bibitem{mp:16}
   M. Park, K. Kalyanam, S. Darbha, P. Khargonekar, M. Pachter, and P. Chandler, \emph{\textquotedblleft Optimal Residential Battery Storage Operations Using Robust Data-Driven Dynamic Programming\textquotedblright,} IEEE Transactions on Automation Science and Engineering, vol. 13, pp. 564-578, 2016.


   \bibitem{ts:21}
   T. Sun and X. Sun, \emph{\textquotedblleft Adaptive Dynamic Programming Scheme for Nonlinear Optimal Control With Unknown Dynamics and Its Application to Turbofan Engines\textquotedblright,} IEEE Transactions on Industrial Informatics", vol. 17, pp. 367-376, 2021.

    \bibitem{bl:19}
   B. Luo, D. Liu, T. Huang, and J. Liu, \emph{\textquotedblleft Tracking Control Based on Adaptive Dynamic Programming With Multistep Policy Evaluation\textquotedblright,} IEEE Transactions on Systems, Man, and Cybernetics: Systems, vol. 49, pp. 2155-2165, 2019.

   
      \bibitem{xs:14}
   X. Xu, C. Lian, L. Zuo, and H. He, \emph{\textquotedblleft Kernel-Based Approximate Dynamic Programming for Real-Time Online Learning Control: An Experimental Study\textquotedblright,} IEEE Transactions on Systems, Man, and Cybernetics: Systems, vol. 22, pp. 146-156, 2014.

     \bibitem{hz:19}    
   H. Zhang, H. Su, K. Zhang, and Y. Luo, \emph{\textquotedblleft Event-Triggered Adaptive Dynamic Programming for Non-Zero-Sum Games of Unknown Nonlinear Systems via Generalized Fuzzy Hyperbolic Models\textquotedblright,} IEEE Transactions on Fuzzy Systems, vol. 27, pp. 2202-2214, 2019.



      \bibitem{dp:05}    
  D. Bertsekas, \emph{\textquotedblleft Dynamic Programming and Optimal Control\textquotedblright,} IEEE Transactions on Fuzzy Systems, Athena Scientific, 2005.


       \bibitem{dp:12}    
   D. Bertsekas, \emph{\textquotedblleft Dynamic Programming and Optimal Control, Vol II: Approximate Dynamic Programming\textquotedblright,} Athena Scientific, 2012.



    \bibitem{lj:11}
   L. Grune and J. Pannek \emph{\textquotedblleft Nonlinear Model Predictive Control
Theory and Algorithms\textquotedblright,} Springer, 2011.



  \bibitem{jl:09}
  J. Liu, D. Pena, and P. Christofides, \emph{\textquotedblleft Distributed Model Predictive Control of Nonlinear Process Systems\textquotedblright,}
  AIChE Journal,
  vol. 55, pp. 1171-1184, 2009.

   \bibitem{mm:20}
M. Maiworm, D. Limon, and R. Findeisen, \emph{\textquotedblleft Online learning-based model predictive control with Gaussian process models and stability guarantees\textquotedblright,}
  International Journal of Robust and Nonlinear Control,
  vol. 31, pp. 8785-8812, 2020.

  



  \bibitem{md:22}
   M. Augustine and D. Patil, \emph{\textquotedblleft A Practically Stabilizing Model Predictive Control Scheme for Switched Affine Systems,\textquotedblright} IEEE Control System Letters, vol. 7, pp. 625-630, Sep. 2022.


    \bibitem{md:21}
   M. Augustine and D. Patil, \emph{\textquotedblleft A Computationally efficient  LQR based Model Predictive Control Scheme for Discrete-time Switched Linear Systems,\textquotedblright}, $60^{th}$ IEEE Conference on Decision and Control, Texas, USA, Dec. 2021


 \bibitem{ind:03}
   Inteco, \emph{\textquotedblleft Multi-Tank User Manual,\textquotedblright} Inteco Ltd, Krakow, 2003.


   \bibitem{wh:12}
W. Zhang, J. Hu and A. Abate,\emph{\textquotedblleft Infinite-Horizon Switched LQR Problems in Discrete Time:  A Suboptimal Algorithm With Performance Analysis,\textquotedblright} 
IEEE Transactions  on Automatic Control, vol.  57, no.  7, pp. 1815-1821, Jul. 2012.


  \bibitem{tg:08}
T. Geyer, G. Papafotiou, and M. Morari,\emph{\textquotedblleft Hybrid model predictive
control of the step-down dc-dc converter,\textquotedblright} 
IEEE Transactions on Control
Systems Technology, vol.  16, pp. 1112-1124,    Nov. 2008. 


 \bibitem{af:20}
A. Forootani, R. Iervolino, M. Tipaldi, and J. Neilson,\emph{\textquotedblleft Approximate Dynamic Programming for Stochastic Resource Allocation Problems,\textquotedblright} 
IEEE/CAA Journal of Automatica Sinica, vol.  7, pp. 975-990,   2020.


 \bibitem{pn:22}
P. Beuchat, J. Warrington, and J. Lygeros,\emph{\textquotedblleft Point Wise Maximum Approach to Approximate Dynamic Programming,\textquotedblright} 
IEEE Transactions  on Automatic Control, vol.  67, pp. 251-266,   2022.






\bibitem{KC1}
K. Chacko, S. Janardhanan, and I. Kar,\emph{\textquotedblleft  Computationally Efficient Nonlinear MPC for Discrete System with Disturbances,\textquotedblright} 
International Journal of Control, Automation and Systems, vol.  15,   2022.



 \bibitem{KC2}
K. Chacko, S. Janardhanan, and I. Kar,\emph{\textquotedblleft Efficient Nonlinear Model Predictive Control for Discrete System with Disturbances,\textquotedblright} 
Int. Conf. on Control, Automation, Robotics and Vision, Singapore,  2018.





\bibitem{sdh:21}
S. Huang, Z. Hu, G. Cao, G Jing, and Y. Liu,\emph{\textquotedblleft  Input-Constrained-Nonlinear-Dynamic-Model-Based Predictive Position Control of Planar Motors,\textquotedblright} 
IEEE Transactions on Industrial Electronics, vol.  68, pp. 50-60,   2021.






\end{thebibliography}

\end{document}